\documentclass[12pt,a4paper]{article}
\usepackage[utf8]{inputenc}
\usepackage{amsmath}
\usepackage{amsfonts}
\usepackage{amssymb}
\usepackage{graphicx}

\providecommand{\keywords}[1]{\textbf{Keywords---} #1}

\usepackage[labelformat=simple]{subcaption}

\usepackage{url}

\usepackage{authblk}
\usepackage{color}
\usepackage{epstopdf}
\usepackage{wrapfig}

\usepackage{pgfplots}
\usepackage{tikz}
\usetikzlibrary{shapes}

\usepackage[
square,
sort,
comma,
numbers
]{natbib}

\usepackage{listings}
\usepackage{color}
\usepackage{xcolor}

\colorlet{punct}{red!60!black}
\definecolor{background}{HTML}{EEEEEE}
\definecolor{delim}{RGB}{20,105,176}
\colorlet{numb}{magenta!60!black}

\definecolor{mygray}{rgb}{0.4,0.4,0.4}
\definecolor{mygreen}{rgb}{0,0.8,0.6}
\definecolor{myorange}{rgb}{1.0,0.4,0}
\definecolor{myblue}{rgb}{0.0,0.4,0.8}

\title{\vspace{-4cm}Towards Blood Flow in the Virtual Human: Efficient Self-Coupling of HemeLB}
\author[1]{McCullough, J.W.S.}
\author[1]{Richardson, R.A.}
\author[1,2]{Patronis, A.}
\author[2]{Halver, R.}
\author[3]{Marshall, R.}
\author[3]{Ruefenacht, M.}
\author[2]{Wylie, B.J.N.}
\author[4]{Odaker, T.}
\author[4]{Wiedemann, M.}
\author[5]{Lloyd, B.}
\author[5]{Neufeld, E.}
\author[2,6]{Sutmann, G.}
\author[3]{Skjellum, A.}
\author[4]{Kranzlm\"{u}ller, D.}
\author[1,7]{Coveney, P.V.}

\affil[1]{Centre for Computational Science, Department of Chemistry, University College London, UK}
\affil[2]{J\"{u}lich Supercomputing Centre, Forschungszentrum J\"{u}lich, Germany}
\affil[3]{SimCenter, University of Tennessee at Chattanooga, USA}
\affil[4]{Leibniz Supercomputing Centre, Leibniz-Rechenzentrum, Germany}
\affil[5]{IT'IS Foundation, Switzerland}
\affil[6]{ICAMS, Ruhr-University Bochum, Germany}
\affil[7]{Informatics Institute, University of Amsterdam, Netherlands}

\date{}

\begin{document}
\maketitle

\begin{abstract}
\emph{[Article accepted for publication in Interface Focus on 23 September 2020 (\url{https://royalsocietypublishing.org/journal/rsfs}); publication is due late 2020.]} Many scientific and medical researchers are working towards the creation of a virtual human - a personalised digital copy of an individual - that will assist in a patient's diagnosis, treatment and recovery. The complex nature of living systems means that the development of this remains a major challenge. We describe progress in enabling the HemeLB lattice Boltzmann code to simulate 3D macroscopic blood flow on a full human scale. Significant developments in memory management and load balancing allow near linear scaling performance of the code on hundreds of thousands of computer cores. Integral to the construction of a virtual human, we also outline the implementation of a self-coupling strategy for HemeLB. This allows simultaneous simulation of arterial and venous vascular trees based on human-specific geometries.
\end{abstract}

\keywords{High Performance Computing, Blood Flow Modelling, Virtual Human, Lattice Boltzmann Method}

\section{Introduction}
The human body is comprised of several complex and interacting subsystems that, in concert, determine its full operation. Each of these depend on mechanisms that span multiple spatial and temporal scales, from sub-cellular processes to directly observable macroscopic properties. The behaviour of these systems is influenced by individual factors such as age, gender, genetics, environment and medical history. All of these must be considered when a patient presents to a clinician for treatment. The development of a virtual human - a digital replica of an individual and their physiological processes - will assist these decisions by allowing multiple courses of treatments to be considered and the optimal one enacted \cite{Kohl2009, Hunter2010, Hunter2013, Chase2016}. \\ 

The transport of blood around the body is integral to physiological function. Vessels transporting blood to and from the heart connect tissue, organs and muscle, providing the oxygen and nutrients needed for their operation. This fundamental nature of the vasculature makes it a pivotal component in the development of a virtual human and is the focus of the present work. The extensive computational and data requirements of modelling a full virtual human will require the resources of next-generation exascale supercomputers. Taking full advantage of these necessitates developing efficient simulation codes and strategies for communication between them on the largest current machines.\\

Many previous studies of large sections of the human vasculature utilise a 1D solver to capture the bloodflow in some or all of the vessels \cite{Sheng1995, Olufsen1999, Qureshi2014, Audebert2017}. Whilst this can be an efficient approach for simulating large, complex networks, it makes many assumptions about the flow behaviour within a vessel. Modelling 3D flow is more computationally demanding but allows high fidelity analysis of flow within all vessels. Coupled 1D-3D models offer a compromise but still do not resolve all features. Full 3D modelling permits local flow features to be identified that are not possible in lower dimensional models, such as wall shear stress distribution throughout the surface of arteries. The use of a 3D model also permits exact simulation of an individual's vasculature. 1D models generally assume that vessels have a circular cross-section that may vary between neighbouring nodes and over-time. Even with patient-specific dimensions, these structural assumptions will lead to homogeneity of solutions between individuals. The use of a 3D model allows simulation results to be exactly constructed for a specific geometry without such assumption.\\

HemeLB \cite{Groen2013, HemeLBweb} is an open-source 3D, lattice Boltzmann based, fluid dynamics solver for the study of blood flow in complex geometries. It has been developed in C++ for, currently, CPU cores only. Unlike many other, more general, open-source lattice Boltzmann codes (e.g.\ Palabos, TCLB, OpenLB and waLBerla \cite{Palabosweb, TCLBweb, OpenLBweb, waLBerlaweb}), HemeLB has been specifically optimised to enable efficient simulation of the sparse geometries that are characteristic of the vascular networks. Since first being published in 2008 \cite{Mazzeo2008}, HemeLB has been used to study a number of different aspects of the cardiovasculature including cerebral flow, retinal flow, vascular remodelling and magnetic drug targetting \cite{Groen2018, Bernabeu2014, Patronis2018}. \\

The development of a virtual human is an ongoing process that will require many computational and algorithmic developments to accurately and efficiently capture physiological behaviour. The purpose of the present paper is to discuss some recent advancements in HemeLB that brings us closer to being able to conduct high-fidelity simulations of the full human vascular tree. In this paper we present proof-of-concept studies that demonstrate the capability of these changes to run a large, 3D flow model on realistic geometries of human-scale arterial and venous trees. These serve as a stepping stone to highlight both the existing capability and the areas which need ongoing development and improvement. The results at this stage are not intended to provide quantitative validation of full-human scale blood flow. This is a target for future work. \\

The paper is structured into three further sections. Section \ref{sec:CompAdv} discusses the computational advancements that have been made within HemeLB to permit the modelling of full human vasculatures. In particular, we discuss the incorporation of next-generation Message Passing Interface (MPI) developments for data communication, the large-scale performance characteristics of HemeLB and the self-coupling of HemeLB. Section \ref{sec:AVCouple} describes the progress that has been made in using the self-coupling of HemeLB to simulate the flow through the arteries and veins of a specific human vasculature. We present results for proof-of-concept models including an illustrative test case and coupled vascular trees of human legs, important steps towards simulation of high fidelity full-human models on exascale machines. The paper concludes in Section \ref{sec:Conclusion} with a discussion of pathways to continue with the current work towards the modelling of a full virtual human.  \\

\section{Computational Advancements}
\label{sec:CompAdv}
HemeLB solves 3D fluid flow using the lattice Boltzmann method. This approach to solving the Navier-Stokes equations is well known to possess excellent parallel computational efficiency, particularly for bulk flow, and is readily adaptable to complex geometries. The current work uses a D3Q19 lattice with a single relaxation time collision operator with pressure and velocity boundary conditions implemented as appropriate. For a more technical description of the core HemeLB implementation we refer the reader to our previous publications \cite{Mazzeo2008, Nash2014, Groen2018}. Here, HemeLB assumes vessel walls to be rigid and characterises the blood as a Newtonian fluid although other rheological models are available \cite{Bernabeu2012,Groen2018}. Although previous HemeLB studies have successfully captured blood flow with these assumptions \cite{Bernabeu2012,Groen2018}, we recognise that as we progress towards the development of a virtual human the physics captured by HemeLB will need ongoing evaluation. For example it is known that the compliance of vessels has an impact on flow behaviour for both arterial and venous networks and can be a factor in cardiovascular disease.\\

In order to successfully model a full virtual human, many physiological and biological processes need to be captured over sufficiently long timescales. In the context of blood flow this pertains to the execution of several cardiac cycles. HemeLB focuses on the macroscopic behaviour of vascular transport and requires  coupling to other codes to model further system interactions. For example, a simulation of the human heart using Alya Red \cite{Vazquez2015}, another high performance computing code, may also be included and coupled to the boundaries of the arterial/venous tree. High-fidelity resolution of the vasculature of a full human requires extremely large quantities of geometry data (billions of data points). Processing such quantities of data and conducting simulations for physically meaningful time scales demands the resources of cutting-edge supercomputers. Here we discuss recent developments of HemeLB that enable this.\\

A drawback of the lattice Boltzmann method is that it stores flow information in a memory intensive manner. HemeLB must initialise and distribute this data within the limits of the available computer. On supercomputers, both of these tasks are non-trivial, meaning that code efficiency is an ever-present development concern. In this section we discuss improvements that have been developed for HemeLB to address these issues and present scaling results to demonstrate their efficacy. Additionally, we outline a framework for the self-coupling of HemeLB that complements these development goals.\\

\subsection{Reduction of Data Communication Within MPI}
One significant challenge in modelling the full human vasculature is generating accurate geometries on which to conduct simulations. Based on studies presented in Section \ref{sec:AVCouple}, at least $10^8$ lattice sites are required in order to simulate flow in the smallest imaged vessels within the same model as large vessels such as the aorta. Efficiently constructing geometries of this magnitude is a significant computational challenge. Attempting to perform this task within the limitations of many supercomputers can cause memory or time restrictions to be exceeded. A particular problem that can be encountered is exceeding the limit on the number of elements that can be communicated within a single MPI operation. We refer to this restriction as the BigCount problem. \\

BigCount is based on the premise that, in the early years of 32-bit computing, the need for an application to track more than a billion items would be rare or too far into the future to be practical.  In computing environments, a single byte using 8 binary bits can express 2$^8$ (or 256) unique values, such as the set of integers from 0 to 255, or from -128 to 127.  If a multi-process application needs to send an array of 1000 items from one process to another, the sending process would need more than one byte to express to the receiving process a count of how many items to expect.  Thus, the generic integer type \verb|int| on a 32-bit system was commonly allotted a width of 4 bytes, even as 64-bit hardware and operating systems were entering the market.  To use more or less than 4 bytes for an integer type, developers using languages like C must often specify explicitly with keywords such as \verb|long| or \verb|double| for the former, or \verb|char| for the latter.  Earlier MPI standards defined counts and displacements to use the generic integer type, which imposed the 32-bit limit on their associated variables. \\

The first steps towards solving BigCount have already been taken by the introduction of a new data type called \verb|MPI_Count| into the MPI-3 standard.  \verb|MPI_Count| can support up to 128-bit integers, though its support in MPI-3 is limited to a few functions \cite{message2012mpi}.  The MPI-4 standard will likely solve BigCount for more commonly-used functions, but its release is still too far into the future for those like us who could benefit from using 64-bit counts today.  Displacements and offsets, values that allow arbitrary access to memory locations in relation to a starting address, are also vulnerable to the 32-bit limit.  The \verb|MPI_Offset| type was added to the current standard, designed to store explicit offsets that may be larger than can be expressed with a 32-bit integer. The \verb|MPI_Aint| type is also in the current standard, and is meant to store memory addresses larger than 4 bytes. \\

A number of transitional measures have been developed to mitigate the BigCount problem until it is solved by the MPI-4 standard.  For example, derived datatypes have been used to express and package large counts into contiguous arrays with the introduction of generalized requests \cite{MPI2}.  Some work has also been done to try to reduce the overhead by offloading the work to graphical processing units \cite{6957231}. Recent discussions among members of the MPI Forum about handling BigCount have focussed on possible solutions, including function pointers, adding `\verb|_x|' variants to standard functions \cite{issue137}, and using  \verb|MPI_Offset| or \verb|MPI_Aint| to handle large displacement values \cite{issue80}. At a user-level, other options for addressing BigCount involve utilising extra software libraries or making modifications to the existing code.  BigMPI is a library that introduces a set of MPI-related functions which accept large counts as \verb|MPI_Count| types that map to MPI-3 calls \cite{Hammond2014}.  BigMPI focuses more on native types over derived types, as its purpose is to support counts that exceed the system-defined maximum (\verb|INT_MAX|) for commonly used functions such as send and receive types, most of the collective operations, and remote memory access functions.  BigMPI converts input stored with large types (and associated data) into a series of smaller supported types (and contiguous data) by way of the MPI standard `\verb|_v|' variants like \verb|scatterv| for collectives. \\

\begin{figure}[h]
  \centering
  \includegraphics[width=\linewidth]{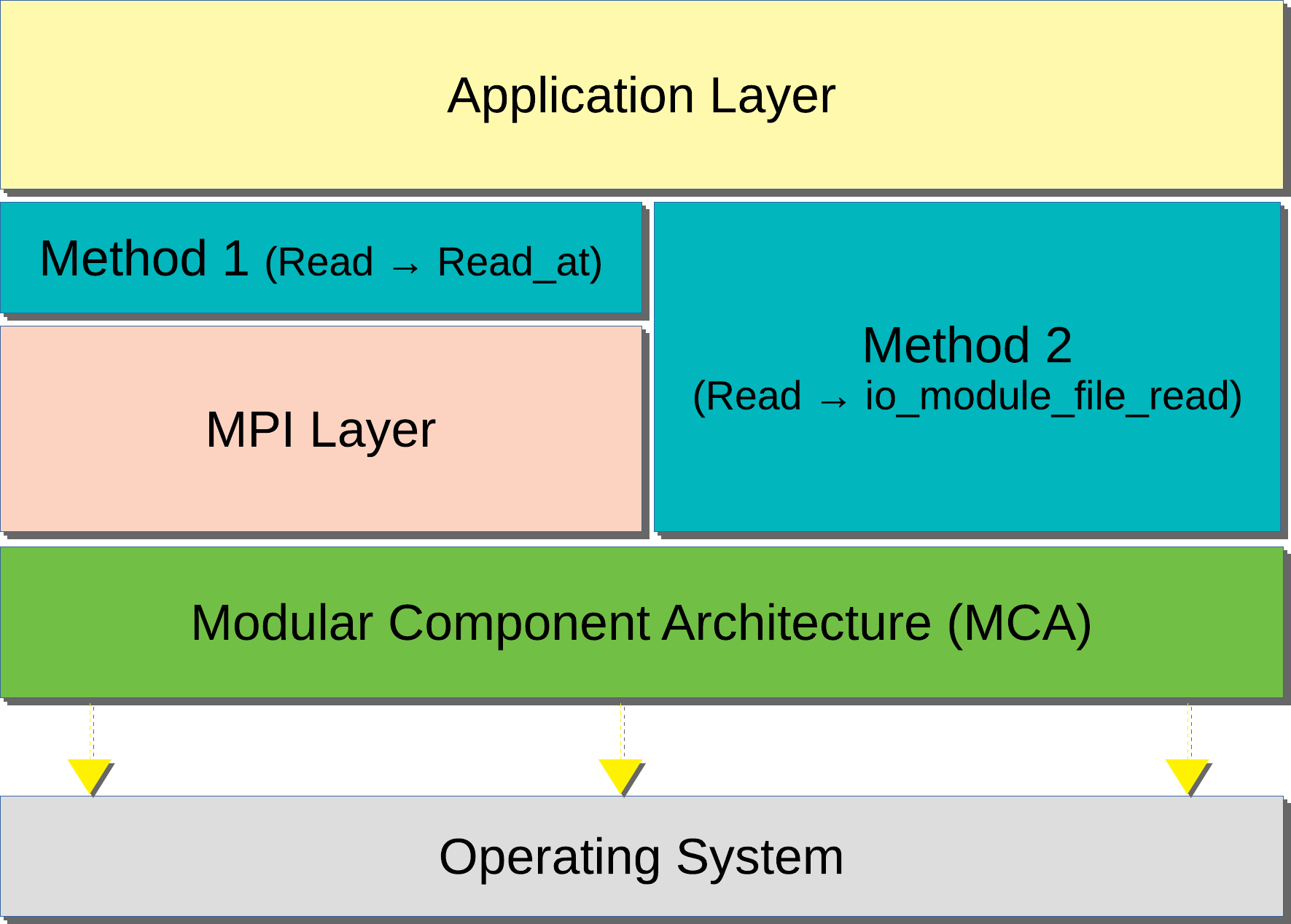}
  \caption{Two methods for working around the 32-bit limitation of MPI communications by breaking large data into segments. Method 1 modifies the existing MPI calls to permit \texttt{MPI\_Count} types. Method 2 bypasses the I/O section to allow usage of the \texttt{MPI\_Count} type.}
\label{layer}
\end{figure}

HemeLB would be able to take advantage of exascale machines today, but is hindered by limitations in the input/output (I/O) component of MPI.  At scale, HemeLB has the opportunity to utilize a large number of data items that exceed the 32-bit limit when calling \verb|MPI_File_read|.  To work around this current limitation, the data is broken into 32-bit safe segments and read in sequence.  Here we describe two methods for achieving this: Method 1, which is functionally identical to application code used in a function called \verb|read_x| (provided in Section 1 of the Supplementary Material), and Method 2, an implementation-specific workaround which bypasses the MPI layer and directly uses an underlying I/O module.  Schematic outlines are presented in Figure \ref{layer}.

\subsubsection{Method 1: Chunking at the MPI API level}

The first method emulates a user space method at the application programming interface (API) level that breaks file reading operations into chunks with manageable count values. We implemented it as a proposed MPI function with the signature below. \\

\begin{lstlisting}[language=C++,firstnumber=1, caption={MPI file read that accepts the MPI\_Count datatype}]
MPI_File_read_x(
  MPI_File fh, void *buf, MPI_Count count,
  MPI_Datatype datatype, MPI_Status *status );
\end{lstlisting}

As shown in Figure \ref{layer}, the new function resides on top of the standard \verb|MPI_File_read|, and accepts a \verb|MPI_Count| type for the item count.  The function will make a series of calls to the standard \verb|MPI_File_read_at|, transparent to the user.  We expect this function to be more convenient for users who regularly deal with large data counts, but cannot modify their MPI implementation.  In most cases, we expect this method to perform marginally faster than \verb|read_x()|, since it makes fewer internal MPI calls compared to the user implementation.   \\

\subsubsection{Method 2: I/O module bypass of the MPI API}

This method is specific to the OpenMPI implementation of MPI, but it could possibly be adapted to other variants.  What makes OpenMPI particularly suitable for Method 2 is its structure, which includes the Modular Component Architecture (MCA).  By default, OpenMPI uses its base I/O module, OMPIO, to read files.  While 32-bit count values are accepted, many of the OMPIO functions could support larger values.  We modified the function signatures through the hierarchy of OMPIO functions to accept the \verb|MPI_Count| type, which works as long as the \verb|INT_MAX| value for the system's C library is large enough to support 64-bit integers. \\

To evaluate the performance of the two methods, we measured the run times of a user space chunking method that avoids large counts against the API layer chunking (Method 1) and the I/O module bypass (Method 2) by reading a series of randomly generated files ranging from 1GB to 32GB in size, and the average time spent in \verb|MPI_File_read_x| over multiple runs. \\

\begin{figure}[!ht]
 \begin{subfigure}{0.98\textwidth}
   \centering
   \includegraphics[width=0.98\textwidth]{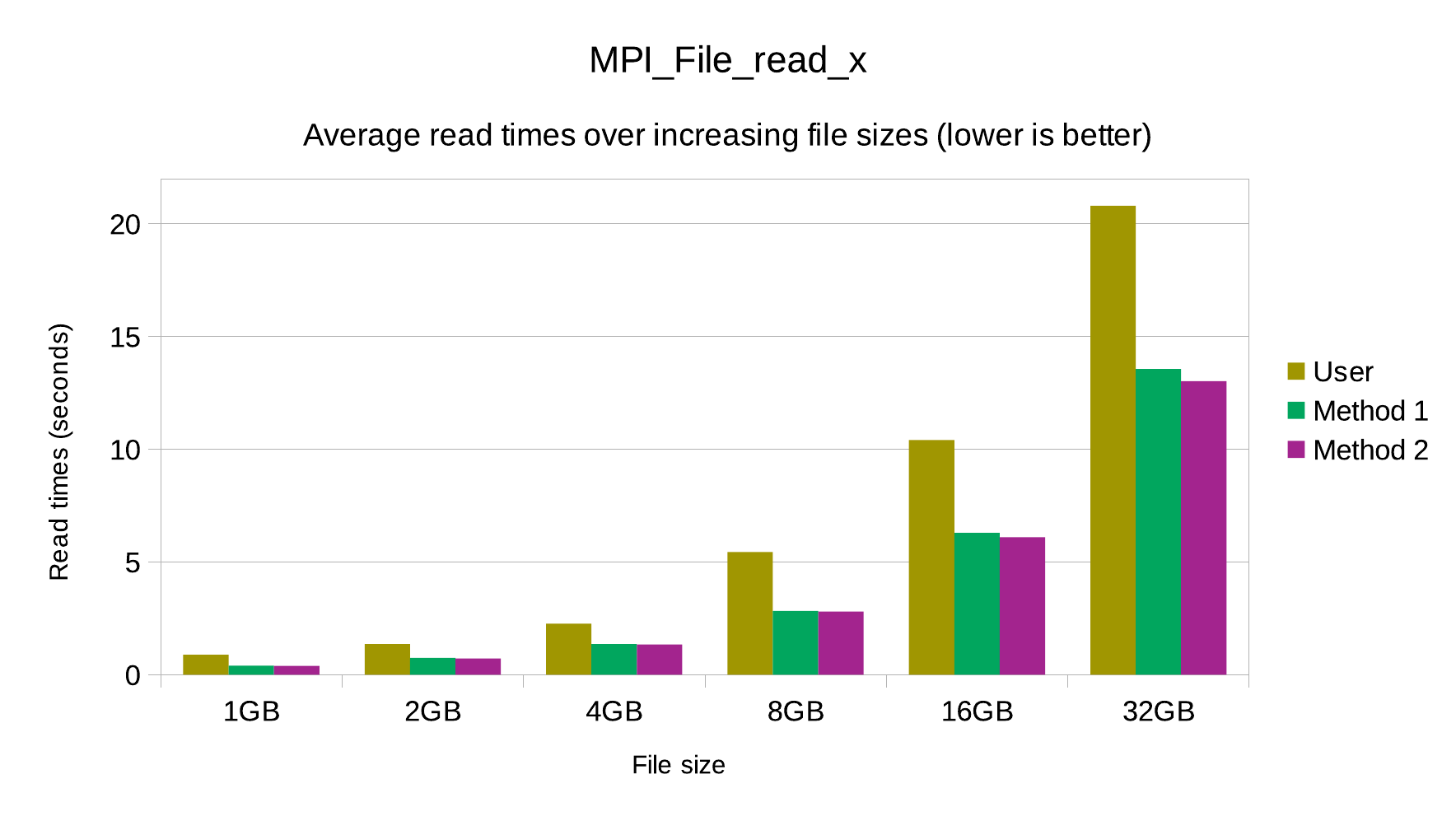}
   \caption{Mean read times}
   \label{readtimes}
 \end{subfigure}
 \begin{subfigure}{0.98\textwidth}
   \centering
   \includegraphics[width=0.98\textwidth]{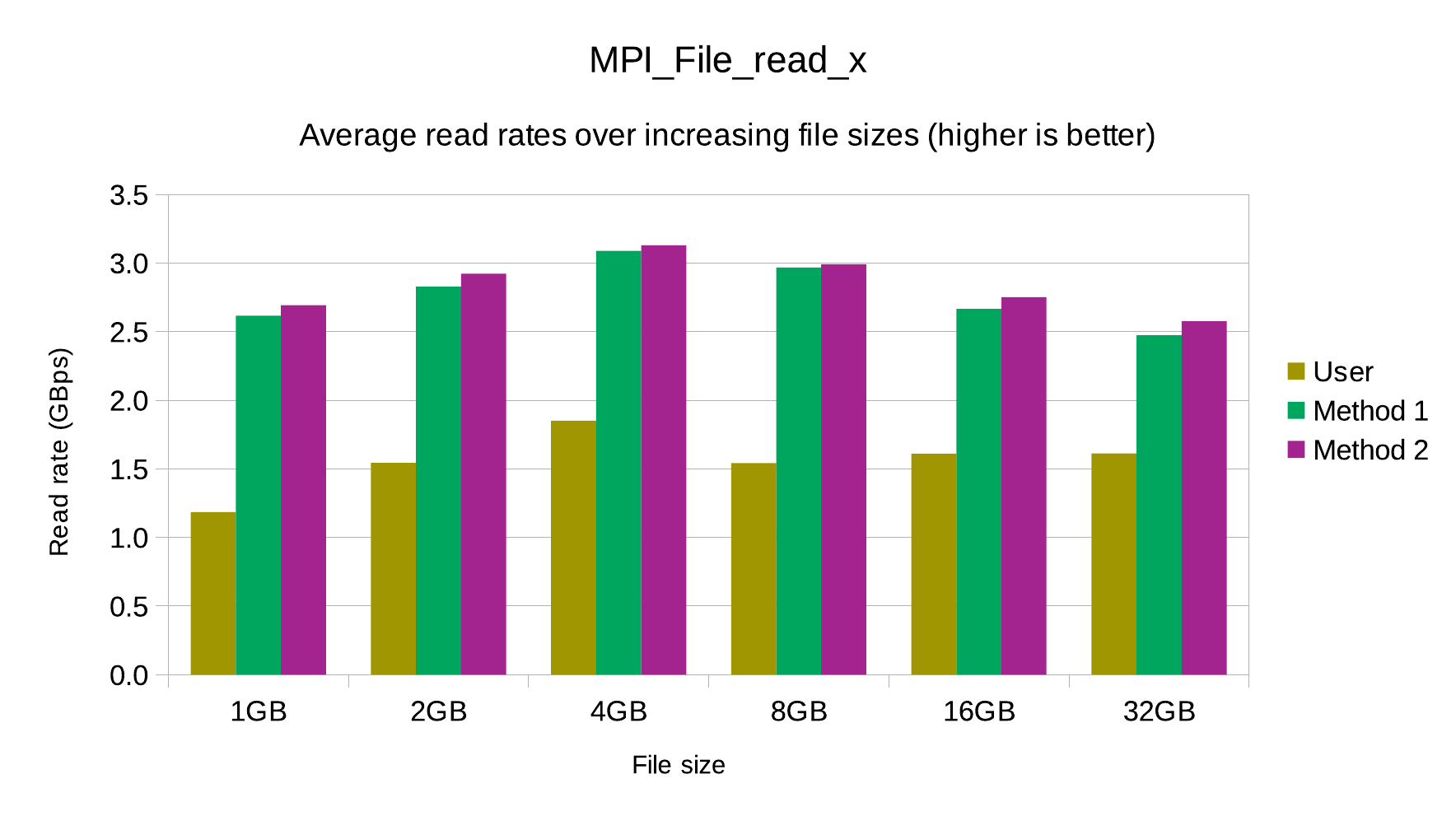}
   \caption{Mean read rates}
   \label{meanreads}
 \end{subfigure}
  \caption{Comparison of the mean read time and read rates observed by user space file chunking versus two different BigCount workarounds in \texttt{MPI\_File\_read\_x} when reading files of various sizes.}
  \label{MeanRatesAndReads}
\end{figure}

Figure \ref{readtimes} shows the time spent in calls to \verb|MPI_File_read_x| as the file sizes increase, in which both methods are observed to be clearly faster than the user space workaround. This advantage becomes more apparent when read rates are compared (Figure \ref{meanreads}). Performance improvements realised with these techniques will enable HemeLB to read large datasets more efficiently.\\

Performance differences aside, the decision to use one method over the other is a practical matter.  Both methods require adding \verb|MPI_File_read_x| to the standard, though Method 1 leaves the rest of the MPI implementation untouched.  Method 2 is specific to OpenMPI and requires slight modification to the underlying I/O module.  In either method, the user only needs to ensure the type for the count is \verb|MPI_Count|. \\

\subsection{Extreme Scale Performance}

HemeLB executions have been audited on different HPC computer systems by the EU Centre of Excellence Performance Optimisation and Productivity (POP)\footnote{\url{https://www.pop-coe.eu/}}.  These performance assessments, based on measurements taken with the highly scalable open-source Scalasca/Score-P toolset~\cite{art:scalasca}, found very good computation and communication efficiencies, while identifying memory consumption and load balance as issues to improve. \\

\begin{figure}
\begin{subfigure}{0.475\textwidth}
 \includegraphics[width=\linewidth]{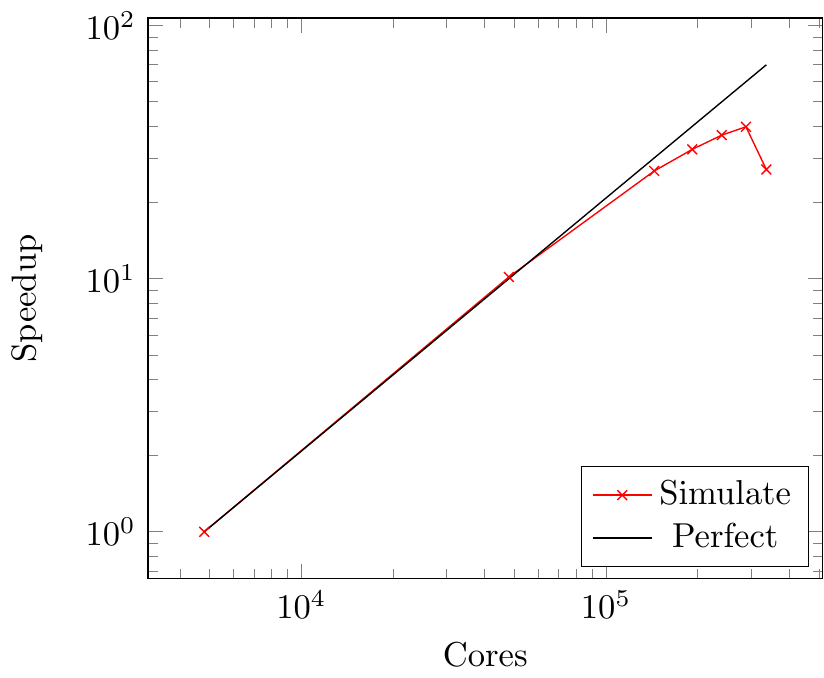}
  \caption{Speed-up relative to 300 nodes (16 cores/node) on Blue Waters.}
  \label{fig:BWscaling}
\end{subfigure}
\hspace{2mm}
\begin{subfigure}{.475\textwidth}
  \includegraphics[width=\columnwidth]{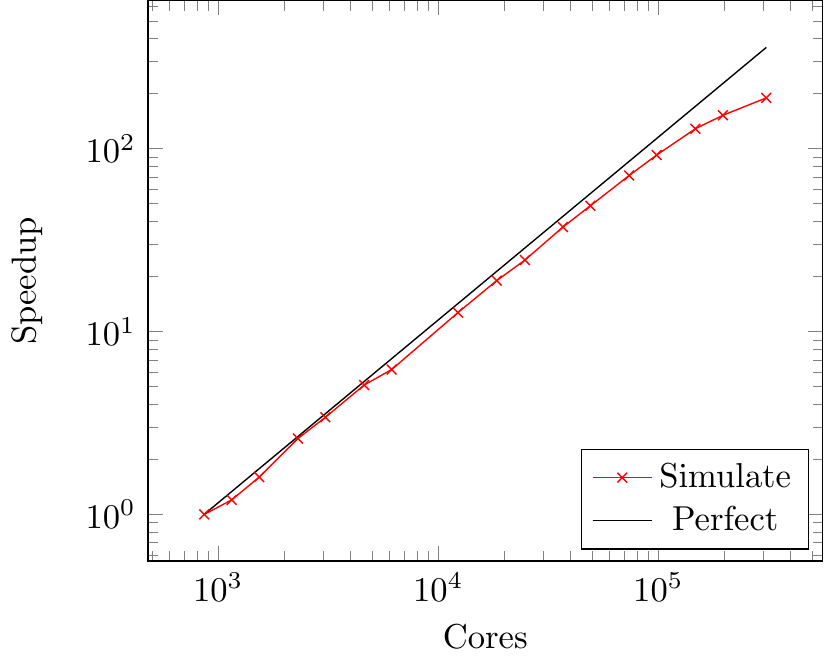}
  \caption{Speed-up relative to 18 nodes (48 cores/node) on SuperMUC-NG.}
  \label{fig:SNGspeedup}
\end{subfigure}
\caption{Scaling of HemeLB simulation time for 5000 lattice update steps of the coW-6.4us.gmy dataset on Blue Waters and SuperMUC-NG. Efficient strong scaling is seen on both systems to very large core counts.}
\label{fig:HemeLBscaling}
\end{figure}

HemeLB has been previously demonstrated to scale exceptionally well up to 100,000 cores \cite{Patronis2018}. We refer the reader to Groen et al. \cite{Groen2013} for a comparison of HemeLB's performance relative to other lattice Boltzmann codes. Furthermore, the code was recently found to scale with good efficiency on 288,000 AMD 6276 Bulldozer-based Interlagos processor cores of 18,000 Cray XE nodes of NCSA Blue Waters (see Figure \ref{fig:BWscaling}). The breakdown in performance at the largest scales on this machine is due to an unfavourable surface-to-volume (communication-to-computation) ratio of partitions (subdomains). Larger models, where computation dominates communication, are required to sustain computational scaling in this regime. In this paper we articulate more recent development on the SuperMUC-NG machine that is based on newer processors with more cores. \\

The SuperMUC-NG supercomputer \cite{SNGinfo} at the Leibniz-Rechenzentrum  (Germany) comprises 6480 compute nodes with dual 24-core Intel Xeon Platinum 8174 @ 3.10GHz (`Skylake') processors.  144 `fat' compute nodes each have 768GB memory, compared to only 96GB memory for the remaining 6336 `thin' compute nodes bundled into eight domains (known as `islands').  The internal interconnect is an Intel OmniPath network, with a fat-tree topology within islands and 1:4 pruned connection between islands.  A high-performance 50TB parallel filesystem is provided by IBM Spectrum Scale (GPFS), with SUSE Linux Enterprise Server (SLES) 12 SP3 operating system. \\

HemeLB was built with the Intel 19.0.4.243 compiler and MPI library. It was configured to use the MPI Shared Memory model within each compute node to reduce memory requirements when loading the initial lattice data.  For scalability testing, a circle of Willis geometry dataset of 21.15GiB was used (similar to that used in \cite{Patronis2018} but with a lattice spacing of approximately 6.4 $\mu$m ). This domain is split into a total of 1,138,236,832 blocks (1376$\times$1087$\times$761), of which 20,740,240 are non-empty. These non-empty blocks contain a total of 10,154,448,502 lattice sites used for simulation. Excluding the initialisation phase and data output, the simulation time for 5,000 update steps of blood flow was recorded to establish a strong scaling benchmark on SuperMUC-NG. \\  

The fat compute nodes were used for executions with 864-6144 MPI processes (18-128 compute nodes), thin nodes were used for larger runs. In all cases 48 MPI processes (one per core) were executed on each compute node. Generally, only a single execution was done at each scale, during regular operation of SuperMUC-NG, with run-to-run variation in the simulation time found to be small (potentially due to variation in allocation of compute nodes and communication interaction with other jobs). \\

Simulation speed-up relative to the smallest execution configuration (18 compute nodes based on memory requirements) for different numbers of compute nodes is plotted in Figure~\ref{fig:SNGspeedup}, along with a comparison to perfect linear scaling. At half machine scale (147,456 cores), a speed-up factor of 128.7 was obtained for a 170.7 factor increase in compute nodes. This corresponds to a scaling efficiency of 75\%. At full machine scale (309,696 cores) the speed-up factor increased to 189.8 reducing simulation time to 83 seconds. \\

The Scalasca/Score-P assessment of the HemeLB results on SuperMUC-NG showed that, overall, very good computational scaling above 87\% is sustained.  Efficient non-blocking communication between neighbouring lattice blocks maintains excellent communication efficiency above 97\%.  The most significant inefficiency at all scales tested is due to the load balance, generally around 80\% but dropping to 72\% in some larger execution configurations.  While this is still fairly good, it presents the largest opportunity for performance improvement and motivated the inclusion of an improved load balancing framework into the HemeLB code. In Section \ref{sec:LoadBalance}, we present initial results for the inclusion of the ALL framework. \\

With 792 compute nodes bundled within island domains, and islands connected via an additional switch, it is notable that no significant simulation performance advantage was observed when inter-domain switches were avoided or reduced.  Small numbers of failed compute nodes throughout SuperMUC-NG can therefore conveniently be avoided, allowing full flexibility in allocating partitions. \\

The HemeLB version used on SuperMUC-NG incorporated optimisations which were essential to be able to set-up and run this size of simulation.  In particular, these included the use of more memory-efficient data structures and MPI shared memory model for each compute node. \\

Hoekstra et al. \cite{Hoekstra2019} observe that the lattice Boltzmann method possesses an algorithmic structure that will enable it to continue its scaling performance on larger supercomputers in the transition to exascale platforms. In part, this is due to the fact that, unlike some algorithms, the lattice Boltzmann method does not possess a hard limit on scalability that inhibits performance at large-scale. This feature assists in enabling HemeLB to study extremely large flow problems. \\

\subsubsection{Load Balancing}
\label{sec:LoadBalance}
When conducting a simulation over hundreds or thousands of compute cores, its parallel efficiency depends strongly on how evenly the workload is distributed. In its default form, HemeLB conducts a basic decomposition that can, when partitioning certain geometries, result in an unbalanced workload between cores. In order to better distribute the HemeLB workload, ongoing investigations have been conducted with specific load-balancing libraries. A combination of 
Zoltan~\cite{ZoltanHomePage,ZoltanIsorropiaOverview2012} and ParMETIS~\cite{METIS} was found to require substantially longer walltimes and more memory due to its extensive pre-processing. Currently, the ALL (A Load-balancing Library) package, developed by the EU E-CAM Centre of Excellence\footnote{\url{https://www.e-cam2020.eu/}}, is being tested with HemeLB. \\

The original load balancing approach successively assigns workload to cores until a nominally even distribution is obtained. Concave geometric domains may force this algorithm to assign non-contiguous regions to the same core, leading to unfavourable communication overhead. Initial studies with ALL have used the package's orthogonal recursive bisection scheme combined with a histogram method. This combination guarantees good load balancing for single connected and convex domains. When compared to the native HemeLB approach, we find both approaches give a balanced workload distribution of $workload_{ave}/workload_{max} \approx 80 \%$ with only a slight dependence on the number of cores used. Inspection shows that the work distribution, based on the prescribed weight of the blocks, is close to being ideally balanced ($>$98\%), which can be observed for the majority of cores. However, the overall performance is limited by a few outliers, which consume about 10-20\% more time. Initial analysis shows that this is not due to communication overhead and also varies between runs with the same process allocation. A possible reason for this behaviour may be found in memory access patterns where memory cannot be allocated contiguously and read/write operations get imbalanced as a result of concurrent memory allocation of processes on the nodes. A more detailed analysis of these results is currently being conducted and this will, ultimately, be used to improve the load balancing of HemeLB on exascale machines. Preliminary results from these studies are presented in Section 2 of the Supplementary Material.

\subsection{Self-coupling of HemeLB}
\begin{figure}[!ht]
\centering
\includegraphics[width=.75\linewidth]{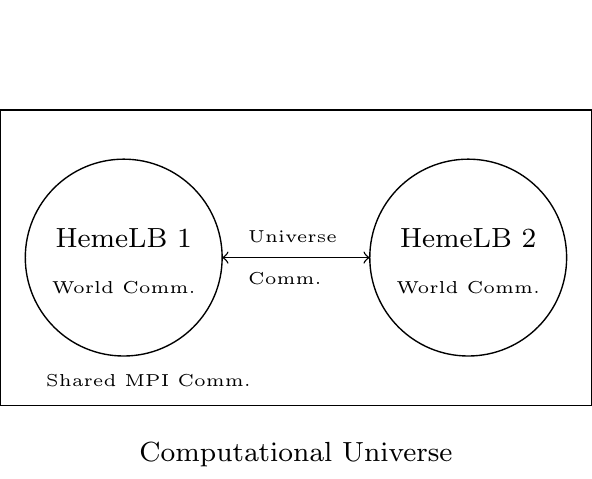}
\caption{Schematic layout of the communications required to couple HemeLB with itself. All communications are conducted using standard MPI calls. This strategy can be extended to permit coupling of further HemeLB worlds. }
\label{fig:CoupledHemeLB}
\end{figure}

In order to create a realistic simulation of a virtual human, HemeLB must be able to communicate with both other HemeLB instances and other biophysical modelling codes. In the first instance, coupling HemeLB to itself will establish a framework for how this can be conducted. Whilst motivated by simulating coupled arterial and venous vasculatures, it also provides a setting to enable simulation of regions at different resolutions. The self-coupled version of HemeLB constructs a common computational universe in which multiple HemeLB instances are instantiated. Each instance retains its own internal communications for executing the simulation on the cores assigned to it. The master ranks of each instance exchange boundary condition states at coupled interfaces to generate interaction between HemeLB worlds. This strategy is laid out schematically in Figure \ref{fig:CoupledHemeLB}. To minimise the data needing to be passed between coupled HemeLB instances, only the average value of macroscopic properties is passed between coupled boundaries. The coupling of the HemeLB instances is carried out by constructing a mapping between the outlets of the first case and the inlets of the second case. In the case of arterial and venous trees there are typically many more venous inlets than arterial outlets, resulting in a number of one-to-many interfaces. Between each of the inlet-outlet pairings, a scaling factor of velocity and pressure is assigned to represent the change in flow behaviour in the intervening vasculature between the two geometries and used to construct local boundary conditions.  

\section{Arterial-Venous Coupling}
\label{sec:AVCouple}
The human circulatory system consists of vessels ranging from the centimetre scale such as the aorta down to micrometre sized capillaries. These are classified (e.g.\ pulmonary/systemic, arteries/veins) based on their transport direction and location. Within a full-human model, the resolution of the smallest vessels and capillaries is often limited. To complete the connection of coupled arterial-venous flows, these vessels are approximated as a sub-scale feature that can be represented with a pressure and velocity drop between vessels on either side. Utilising this approach, the self-coupling of HemeLB will enable simultaneous blood flow simulations of arterial and venous networks. If found to be necessary, future work may involve developing and implementing more sophisticated coupling strategies to represent the sub-scale material, such as through incorporation of porous media models of capillary beds. \\

\begin{figure}
\centering
\includegraphics[width=0.75\textwidth]{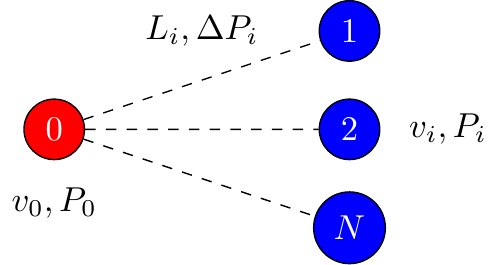}
\caption{A one-to-many coupling is generated to link an arterial outlet (red) with venous (blue) inlets. Each boundary site has an associated velocity, $v$ and pressure, $P$. The distance between each vein and artery, $L_i$, is measured and the associated pressure drop $\Delta P_i$ is defined.}
\label{fig:AVnetwork}
\end{figure}

To simulate coupled vascular networks, a strategy for determining the pressure and velocity scaling factors between coupled boundary locations was developed to ensure that mass conservation is maintained. For a given vessel size there are, generally, many more veins than arteries. In the vascular networks simulated in this study (described later in this section) there are 13 times more vein inlets than arterial outlets. The first stage in defining a coupling map between arteries and veins involves identifying which inlets and outlets of the independent geometries are linked. A na\"ive initial approach is based on the relative proximity of boundary locations. A mapping algorithm was devised to ensure that each arterial outlet is coupled to at least one vein. For each artery, the list of venous inlets is inspected and the closest is assigned to that vessel then removed from the list. Once complete, the remaining veins are assigned to their closest artery. This generates a 1:N coupling map for each arterial outlet, as shown in Figure \ref{fig:AVnetwork}. The distance between each coupled pair is stored for later use. \\

To represent the sub-scale material between the macroscopic arteries and veins a pressure drop between each inlet-outlet pair is assigned to a random value between 0.3 and 0.7. This range represents physiologically expected values \cite{AandPtext2016} and permits variability of structure in the hypothetical capillary network it represents. This correlates directly to the pressure scale factor, $\Delta P_i / P_0$, assigned to that pairing. The velocity scale factor, $v_i / v_0$, is determined by assuming that for each inlet-outlet pair the pressure drop, $\Delta P_i$, is proportional to the product of flow rate, $q_i$, and length, $L_i$,

\begin{equation}
\Delta P_i = P_0 - P_i = k_i q_i L_i.
\label{eq:PressureFactor}
\end{equation}

\noindent By comparison of units, it can be noted that $k_i$ is inversely proportional to the area of the inlet, $A_i$. The scaling factor for velocity is then determined by combining Equation \ref{eq:PressureFactor} for each inlet $i$ with mass conservation of the system (i.e.\ $q_0 = \sum\limits^N_{i=1} q_i$) to determine the scaling of velocity, $v_i$, for each inlet-outlet pair:

\begin{equation}
\dfrac{v_i}{v_0} = \dfrac{A_0}{A_i}\dfrac{\Delta P_iL_1A_i}{\Delta P_1L_iA_1}\left(\sum\limits^N_{j=1}\dfrac{\Delta P_jL_1A_j}{\Delta P_1L_jA_1}\right)^{-1}.
\end{equation}

\noindent Note that a subscript of zero indicates values relating to the arterial outlet. \\

These factors are used to construct the velocity boundary conditions on the opposite geometry. In the `forward' (i.e.\ arterial-to-venous) direction the scale factors are applied to the average velocity at the arterial outlet to construct boundary condition values to be applied at the coupled venous inlets. In the `reverse' direction (i.e.\ venous-to-arterial), the boundary force values at the arterial outlets are determined by taking the average of the average velocities at the inlets scaled by the appropriate scaling factor. To represent the volume of fluid existing in the capillaries between any given coupled inlet and outlet, an explicit force is applied to the outlet side based on the dynamic pressure at that location using the approach of Guo et al. \cite{Guo2002}. The coupling strategy alternates between the `forward' and `reverse' directions each time boundary information is swapped.  \\

\begin{figure}
\begin{subfigure}{0.48\textwidth}
\centering
\includegraphics[width=0.5\textwidth]{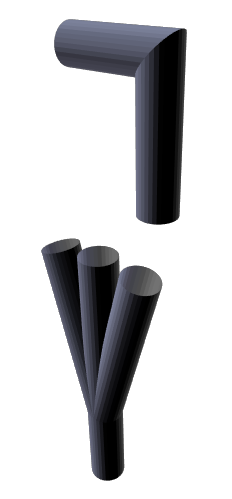}
\caption{Test layout}
\label{fig:OneManyLayout}
\end{subfigure}
\begin{subfigure}{0.48\textwidth}
\centering
\includegraphics[width=\textwidth]{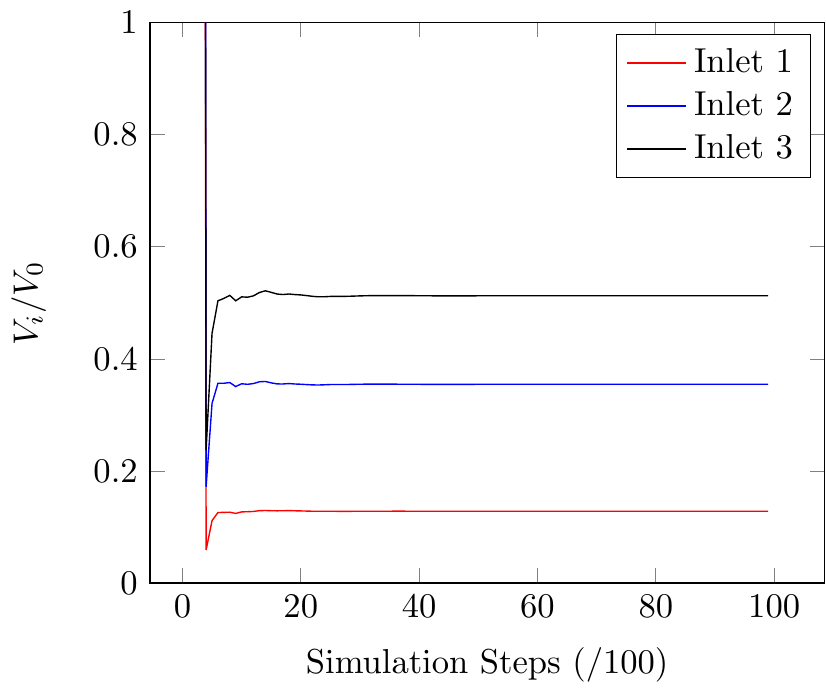}
\caption{Velocity ratios}
\label{fig:OneManyResults}
\end{subfigure}
\caption{Schematic layout and velocity ratio results of the 1:3 test domain for demonstrating performance of the self-coupling of HemeLB, arteries are upper section of the layout.}
\label{fig:OneManyLayoutResults}
\end{figure}

As a demonstration of the performance of the coupling strategy, we show the recorded velocity ratios at coupled inlets and outlets in a simplified 1:3 geometry. This is illustrated in Figure \ref{fig:OneManyLayout}. In this setting the imposed pressure scaling factors were $\Delta P_i/P_0$ = 0.30, 0.52, 0.69 for the $i=1,2,3$ inlets; the corresponding velocity factors were $v_i/v_0$ = 0.13, 0.36, 0.52 with all boundaries of equal size. Figure \ref{fig:OneManyResults} demonstrates that these ratios are achieved between the appropriate inlets when steady-state is reached (again, after approximately 3000 simulation steps). These results verify the boundary condition implementation by demonstrating that the desired results are achieved.\\

\subsection{Obtaining Human-scale Input Data}
\begin{figure}[!ht]
\centering
\begin{subfigure}{.24\textwidth}
  \centering
  \includegraphics[width=\linewidth]{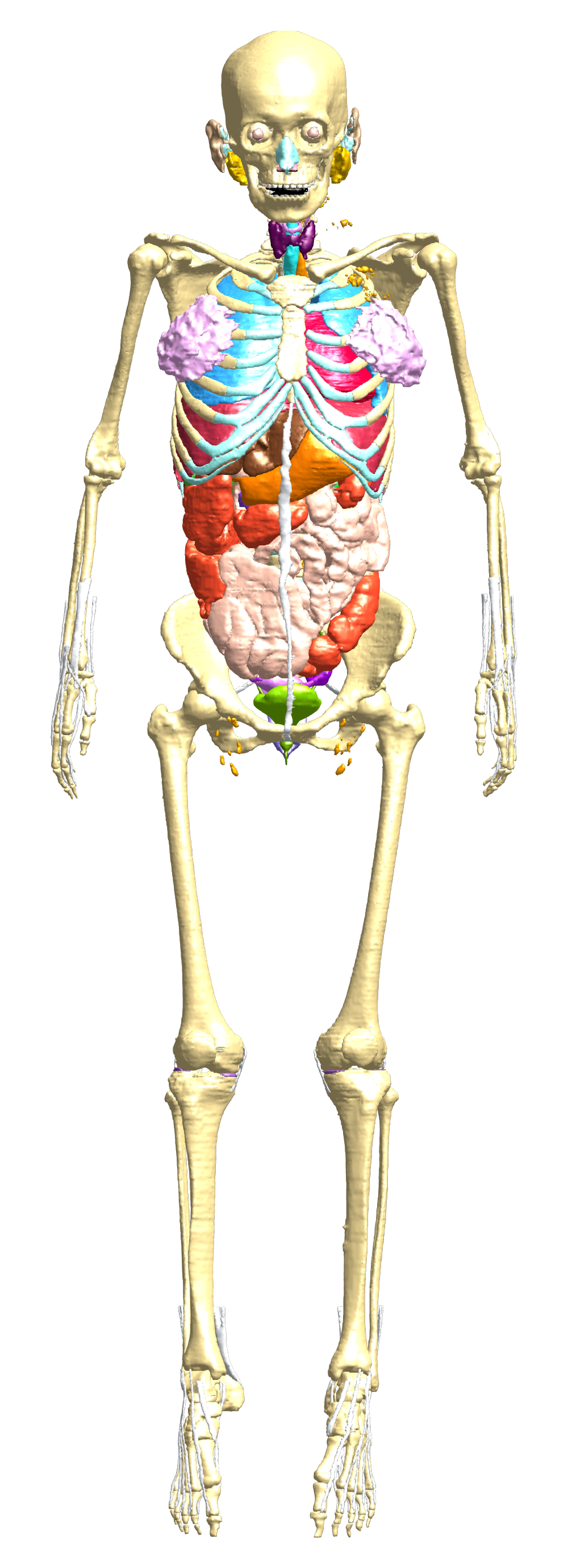}
\end{subfigure}%
\begin{subfigure}{.24\textwidth}
  \centering
  \includegraphics[width=\linewidth]{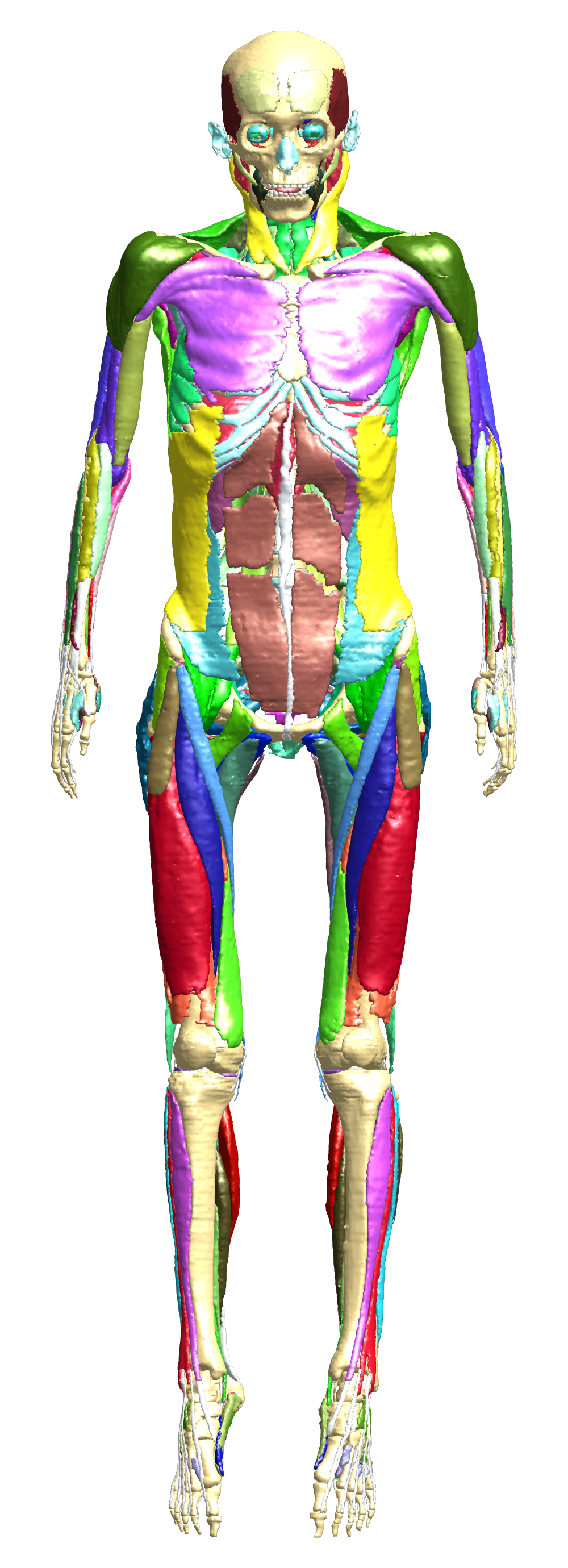}
\end{subfigure}%
\begin{subfigure}{.24\textwidth}
  \centering
  \includegraphics[width=\linewidth]{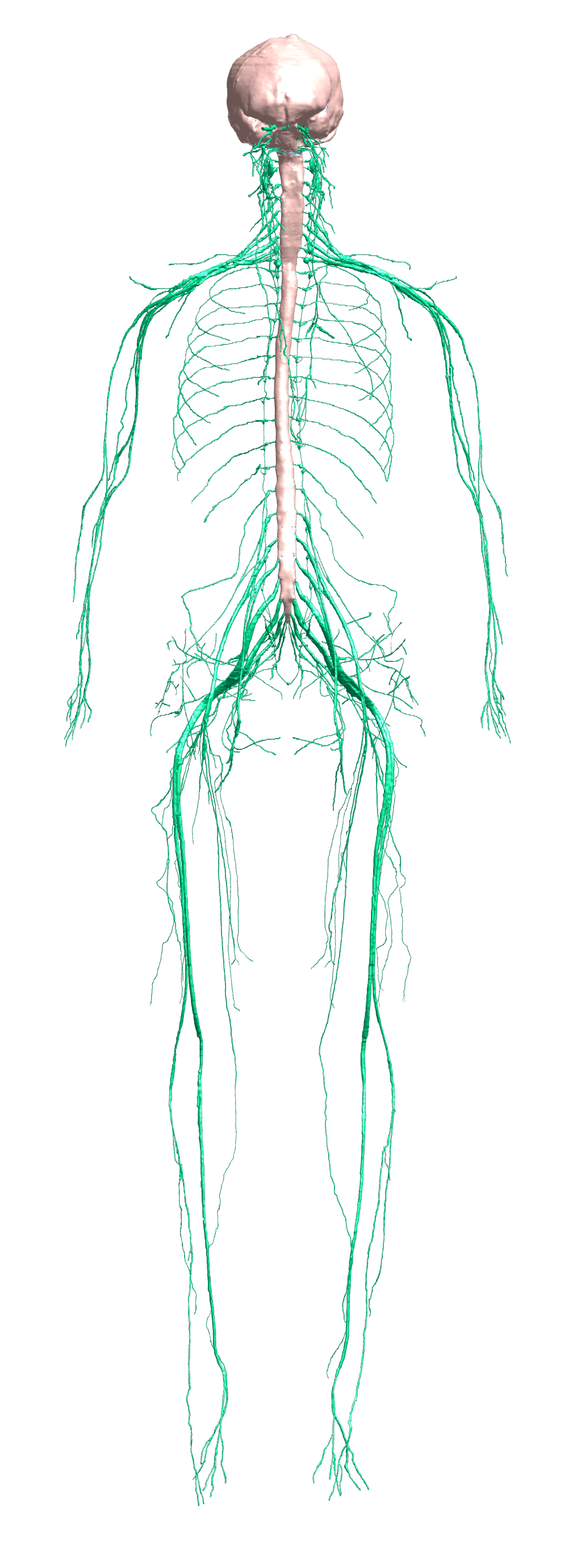}
\end{subfigure}
\begin{subfigure}{.24\textwidth}
  \centering
  \includegraphics[width=\linewidth]{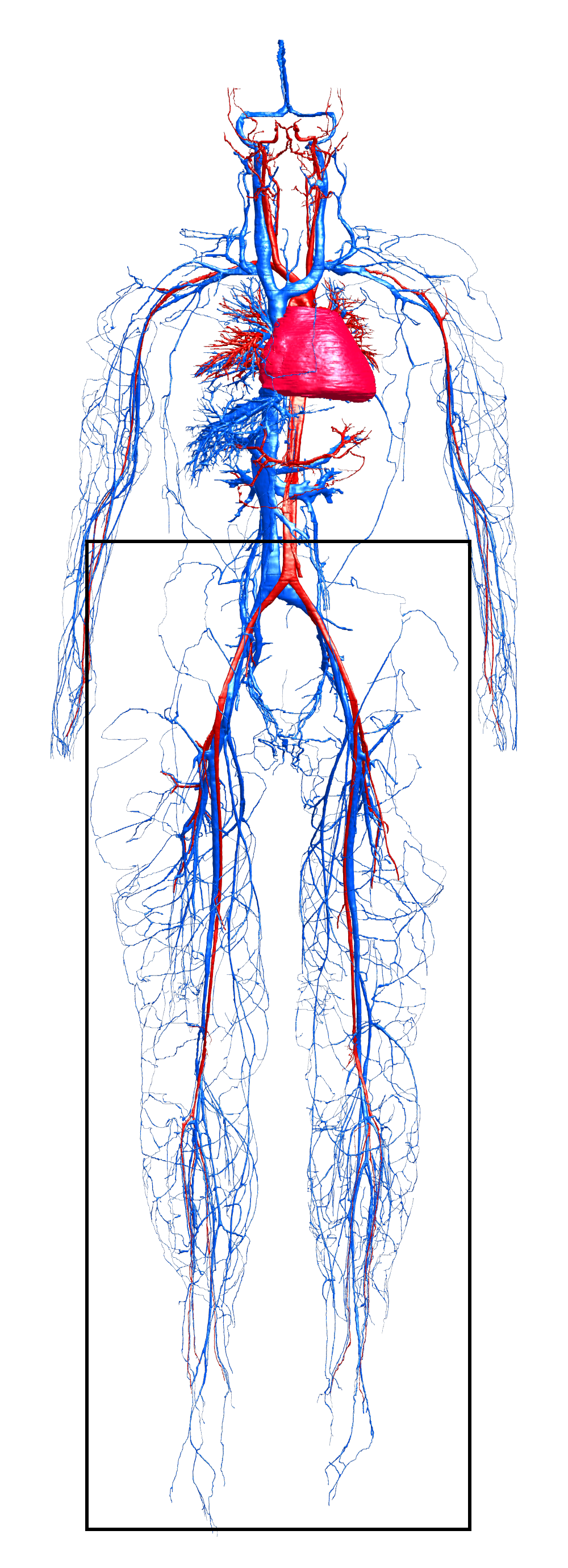}
\end{subfigure}
\caption{The female ViP model Yoon-sun with unprecedented detail of the (from left to right) skeleton and internal organs, muscles, peripheral nerves and vasculature \cite{Park2005,Christ2010, Gosselin2014}. Colours represent different structures in each case. The section of leg vessels highlighted by the box was used for initial tests of human-scale vasculature simulations.}
\label{fig:VHmodels}
\end{figure}

We used the recently created computational anatomical model Yoon-sun \cite{ITIS2019}, which was segmented from high-resolution (0.1$\times$0.1$\times$0.2 mm) colour photographs of cross-sections from a frozen female cadaver obtained in the Visible Korean project \cite{Park2005}. Yoon-sun, available as part of the Virtual Population (ViP) library \cite{Christ2010, Gosselin2014}, was segmented at 0.2$\times$0.2$\times$1.0 mm resolution, separating more than 1100 tissues, and provides unprecedented details in the peripheral nervous system, muscles and arterial-venous system (Figure \ref{fig:VHmodels}). Yoon-sun was created following standardized quality assurance guidelines developed to ensure consistent segmentations across the ViP library. The finest segmented vessels have a diameter in the range of 1-2 pixels (0.2-0.4mm). \\

\begin{figure}[!ht]
\centering
\includegraphics[width=\textwidth]{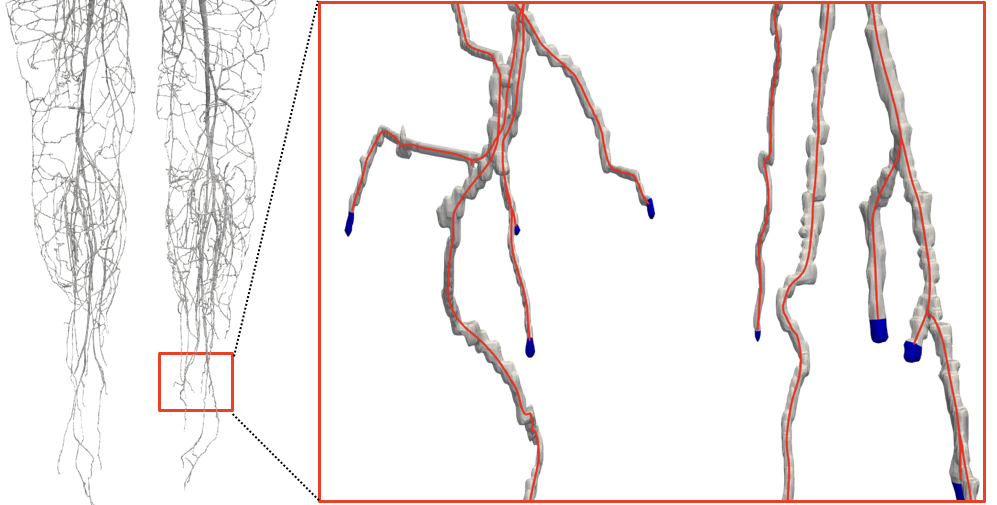}
\caption{Illustration of vessel surface pre-processing for HemeLB, which detects locations for boundary conditions at open edges of the vessel surface. Centrelines (red lines) are generated from the triangulated surfaces of arterial and venous trees. The vessel tips (depicted in blue) are clipped by a cutting algorithm.   }
\label{fig:cut_vessels}
\end{figure}

The surface model of the arteries and veins was clipped at the vessel tips to improve the ability of HemeLB to apply boundary conditions to the geometry. To handle the large number of vessel tips (more than 1500) this was automated through the following algorithm: i) surfaces of the vasculature were extracted using the Marching Cubes algorithm \cite{Lorensen1987};  ii) centrelines were extracted from this triangulated surface using a skeletonisation technique \cite{Tagliasacchi2012, Fabri2011}; and, finally, iii) the vasculature model was clipped by cutting the surface mesh at the end points of the centrelines (perpendicular to the line orientation). This algorithm is implemented using in-house C++ building-blocks, except for the skeletonisation which uses the CGAL library \cite{Fabri2011}. To limit the cut to the corresponding tip surface (i.e.\ to avoid cutting with an infinite plane), a traversal algorithm was implemented, which starts the cut at the vertex closest to the tip location. Figure \ref{fig:cut_vessels} shows an example from the arterial tree. \\

\subsection{Visualisation of Very Large Datasets}
Additionally, we are developing a visualisation method for the analysis of data and communication of results produced by HemeLB. Previous work with HemeLB datasets - developed in tandem with the `Virtual Humans' IMAX movie \cite{VirtualHumanFIlm2018} - focussed on simulating movement of particles based on a much smaller dataset (about 160 million data sites/time step) than is required for our virtual human-scale simulation. This earlier work, using sparsely populated octrees, was used for real-time visualisation and data processing for cinematic renderings \cite{VirtualHumanFIlm2018}.  However, this approach is no longer viable due to the size and complexity of datasets planned to be generated with the self-coupled HemeLB code. Instead, we aim to provide precomputed flow data to visualise flow in complex, large-scale datasets. Whilst this limits versatility of the visualisation it allows for real-time exploration of the flow data that can be directed and altered to investigate different scenarios. \\

\subsection{Human-scale Blood Flow Results}
\begin{figure}[!ht]
\centering
\begin{subfigure}{.75\textwidth}
  \centering
  \includegraphics[width=\linewidth]{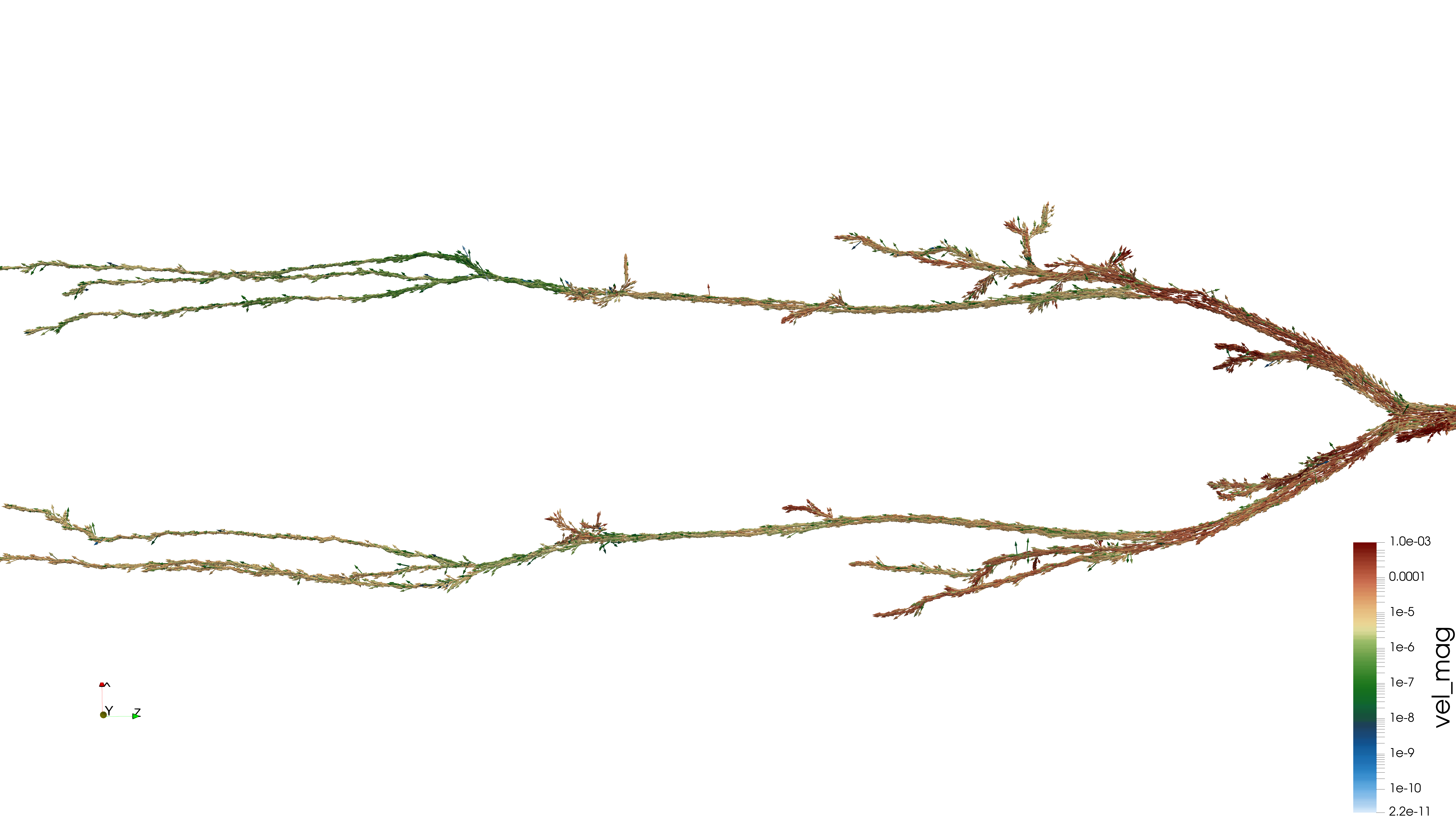}
\end{subfigure}
\begin{subfigure}{.75\textwidth}
  \centering
  \includegraphics[width=\linewidth]{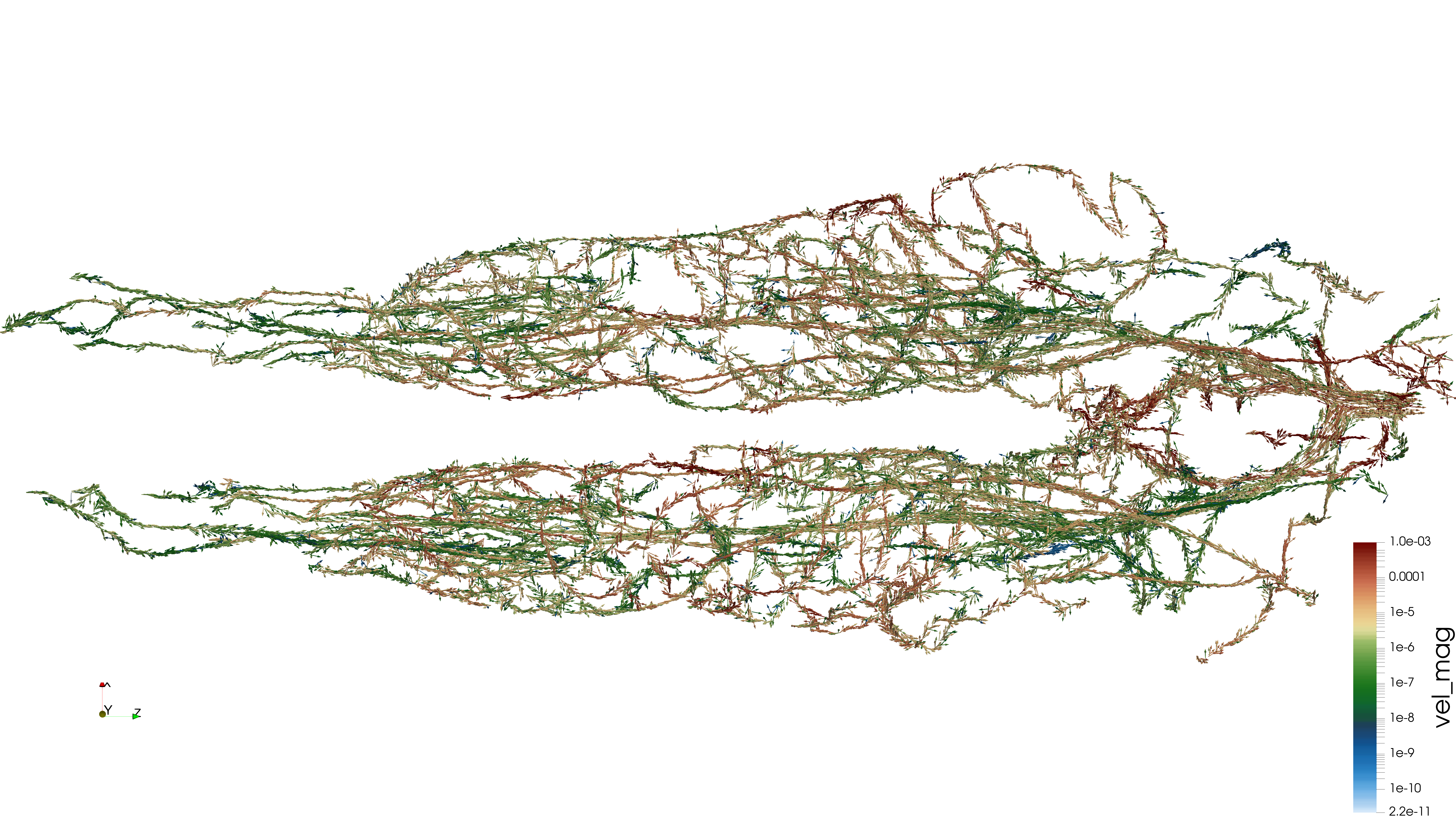}
\end{subfigure}
\begin{subfigure}{.48\textwidth}
  \centering
  \includegraphics[width=\linewidth]{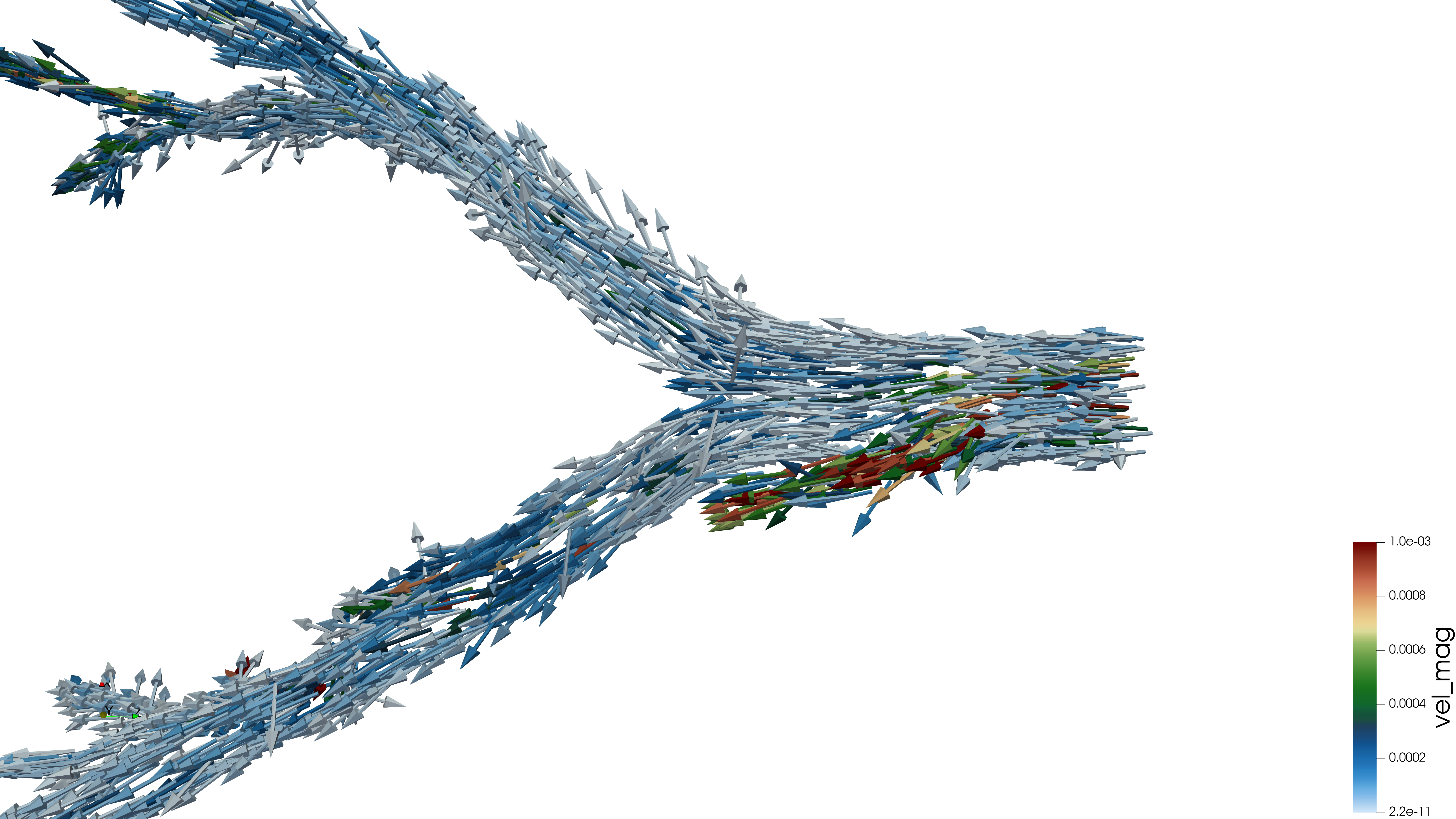}
\end{subfigure}
\begin{subfigure}{.48\textwidth}
  \centering
  \includegraphics[width=\linewidth]{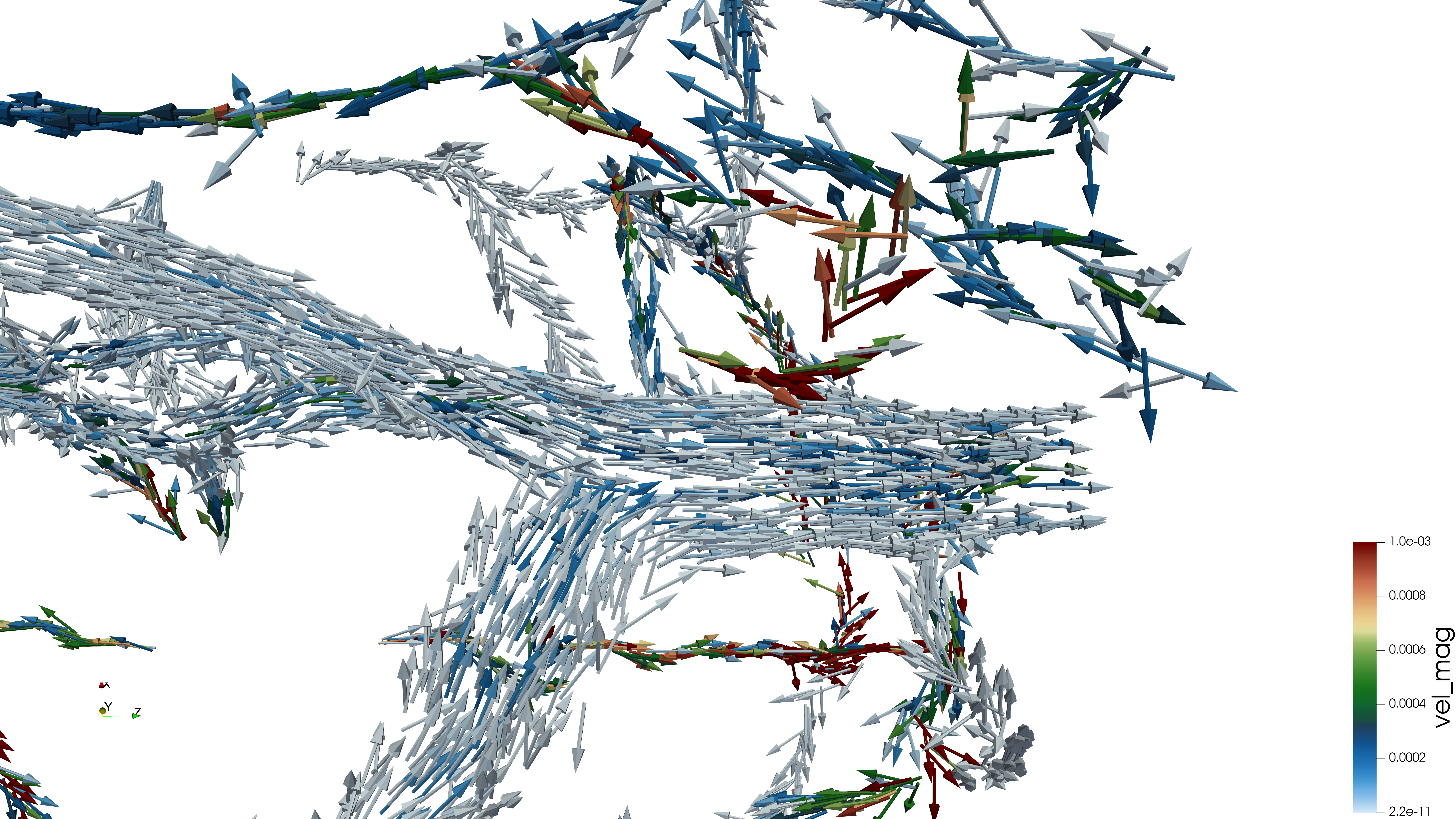}
\end{subfigure}

\caption{Field of velocity vectors (coloured by magnitude on a logarithmic scale) through the arterial (upper row) and venous (centre row) domains. The lower row (arterial left and venous right) is a close-up view of the iliac bifurcation of the aorta at the top of the simulated geometry (right-hand side of full domain images). All figures have been capped at the arterial venous velocity to illustrate the progression of flow through the domain at this point of the simulation. The full domain of the venous tree contains approximately 700,000 vectors. These figures demonstrate the scale of 3D vascular models which we can currently simulate with HemeLB.}
\label{fig:AVvisualisation}
\end{figure}

We simulate blood flow from just above the iliac bifurcation of the Yoon-sun geometry, as indicated in Figure \ref{fig:VHmodels}. In this configuration, the arterial tree has a single inlet and 38 outlets whilst the venous tree has 494 inlets and one outlet. For this case study, the domain was resolved with a lattice spacing of 75 $\mu$m (total of 434,579,134 lattice sites across the arterial and venous geometries). For more detailed simulations a resolution of perhaps 25 $\mu$m would be necessary to fully resolve the finest vessels. Models of this scale would exceed $10^{9}$ lattice sites. Velocity conditions were applied at the inlets to domains and pressure boundary conditions applied to the outlets. The maximum velocity at the inlet to the arterial tree was set at a constant 0.001m/s, smaller than physically expected at that location but necessary to simulate stable flow for this resolution model. This simulation is intended to provide a proof-of-concept of the self-coupling strategy outlined in this paper on human-scale vascular geometries rather than a rigorous quantitative evaluation of its performance. The flow conditions chosen are adequate for the current demonstration purpose. Full validation of physiologically accurate flow will be the subject of future work. The parameters applied to the coupled boundary locations were determined by the self-coupling strategy described in this paper. The simulations used to generate this data were run for 1,000,000 steps, corresponding to 100 seconds of physical time. This took approximately 13 hours on 10,080 cores of SuperMUC-NG. The resulting velocity distribution (Figure \ref{fig:AVvisualisation}) illustrates how the flow development is occurring throughout the full geometry and around the iliac bifurcation. The flow field is, qualitatively, as expected throughout the domain. One issue with conducting simulations on this scale and geometry is the time taken for flow to develop throughout the domain. This highlights the need to carefully consider the initialisation state of large domains so as to rapidly develop physiologically realistic states that minimise simulation time. \\

\begin{figure}[!ht]
\centering
\includegraphics[width=0.75\textwidth]{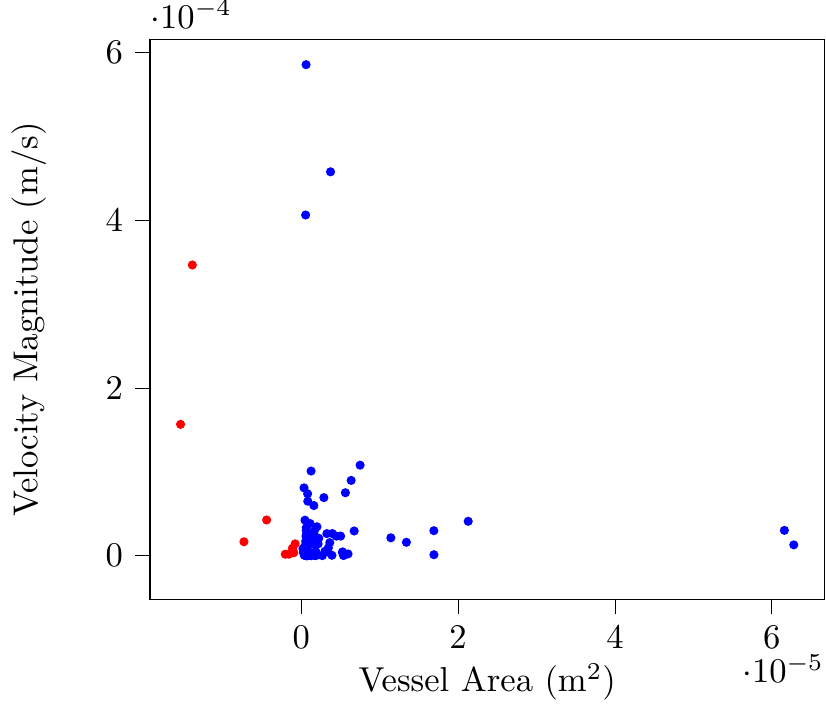}
\caption{Plot of flow velocity magnitude against vessel area after 1,000,000 simulation steps for the coupled arterial (red) and venous (blue) networks of the region indicated in Figure \ref{fig:VHmodels}. Planes of vessels were extracted at 3 locations along the length of the domain. The `negative' region for the arterial vessels in an artefact designed to keep the arterial and venous points separated on the graph.}
\label{fig:PvAplot}
\end{figure}

To provide further qualitative validation of the simulation we examine the flow velocity experienced in vessels compared to their size. Within the systemic circulatory system, the mean velocity is maximal upon leaving the heart and reduces as it travels through to smaller vessels. Within the venous system, the velocity increases again on returning to the heart but at a more gradual rate. Schematic illustrations of this are provided in Chapter 20.2 of \cite{AandPtext2016}. Figure \ref{fig:PvAplot} illustrates the distribution of velocity magnitude as a function of vessel area. Note that the plot is organised from left-to-right based on general progression through the circulatory system; arteries are plotted with a `negative' area. These data points were obtained by inserting a measurement plane at approximately one-quarter, one-half and three-quarters of the way along the length of the geometry. These were separated into cross-sections of the various vessels and mean velocity calculated. The area was estimated by taking a convex hull around the lattice points associated with each vessel. The anticipated velocity reduction in the arterial vessels is clearly observed. The results for the venous tree also generally demonstrate the correct trend - an increase in velocity with vessel area but at a much slower rate-of-change than that seen in the arteries. A small number of higher velocity data points are still present in the venous tree, particularly for smaller vessels. These may arise due to an inappropriate coupling being generated by the na\"ive and automated initial strategy. This indicates that the coupling of complex vascular networks may require a more subtle strategy than the solely distance based approach outlined here  as an initial proof-of-concept. It is also possible that these high velocity data points are driving instabilities that resulted in a low inlet velocity being required in this simulation. The assumption of rigid walls may also impact venous behaviour more strongly than the arterial tree. Addressing these will be a focus for future work.\\

The accuracy and stability of single relaxation time lattice Boltzmann simulations are governed by two equations. The first, $ \nu = \frac{1}{3} \left( \tau -\frac{1}{2} \right) \frac{\Delta x^2}{\Delta t} $,  links fluid viscosity, $\nu$, with the grid spacing, $\Delta x$, time step, $\Delta t$, and relaxation time, $\tau$. The second, $Ma = \frac{\sqrt{3}v_{max} \Delta t}{\Delta x} < 0.1$, stipulates the valid computational Mach number for a simulation. Typically it is desired that $\tau \in \left(0.5,1\right]$ and preferably closer to one; and the Mach number to be as small as possible. These are competing objectives. Rough calculations suggest that for physiologically realistic flows (e.g.\ 1m/s in the aorta) a resolution of 25$\mu$m, and a time step of 1$\mu$s represents the cusp of achievable simulation parameters. Randles et al. \cite{Randles2015} have generated an arterial geometry with over 10$^{11}$ sites at 10 $\mu$m, indicating that this is achievable. These values represent a reason for using large supercomputers to conduct these simulations. Matching of Reynolds numbers with increased fluid viscosity could assist in further relaxing these requirements. \\

Once communication between the coupled instances of HemeLB has been optimised, the full performance profile of the coupling strategy can be assessed. This will require a sufficiently large pair of geometries to study this adequately, perhaps requiring at least 10$^9$ fluid sites each. As demonstrated in Section 2.2, HemeLB has excellent strong scaling performance as a single instance, although the coupling of multiple instances reduces this behaviour due to additional communication overhead. We plan to assess this behaviour of self-coupled HemeLB in a future publication.

\section{Conclusion}
\label{sec:Conclusion}

Significant scientific and medical research effort is being directed towards enabling the simulation of a virtual human. The ultimate delivery of this goal will allow for the prescription of medical treatments personalised for every patient. The virtual human would also allow healthy individuals to discover how they can improve their current lifestyle choices. The performance of supercomputers is now reaching a level at which the creation of a virtual human is feasible. This paper brings together cutting-edge technologies and algorithms for building  and simulating virtual humans on the latest high-performance computing platforms. Firstly, we described software implementations that permit strong scaling performance for complex domains with $\mathcal{O}(10^{10})$ lattice sites on $\mathcal{O}(10^{5})$ cores. In particular, this has involved modifying MPI communication behaviour to allow for larger data communication. Achieving performance on this scale also requires streamlining of internal HemeLB data structures and improved load balancing techniques. We have outlined a self-coupling strategy to allow coupled simulations of arterial and venous flows (a critical feature for modelling blood flow within a virtual human). Here we demonstrate that the coupling strategy is able to communicate information between several hundred linked boundary locations and is able to reconstruct flow at this scale. \\

In this paper, we have demonstrated that the self-coupling of HemeLB is able to simulate blood flow in human-scale geometries that capture the expected dynamic flow features. In future developments of this work, we plan to progress towards creating a virtual human by extending the current simulation domain to the full human and continuing to expand the physics modelled by HemeLB. In particular, we will look at methods for improving the coupling behaviour between domains and relaxing HemeLB's structural and fluid assumptions. In future work, we plan to validate the coupling model presented here by close comparison with physiological data. This will allow us to apply the model to large-scale physiological and clinical situations. Visualisation remains an essential element in comprehension and communication of the large-scale data generated by simulations of human-scale domains. We seek to further develop techniques for the visualisation of human-scale blood flow in order to better enable understanding of the virtual human. The advent of exascale computers will greatly facilitate these endeavours. As many future exascale machines are planned to incorporate the use of accelerators, we are also developing a GPU-enabled version of HemeLB to further capitalise on the performance capabilities of these new machines.\\

\section*{Funding}
We acknowledge funding support from European Commission CompBioMed Centre of Excellence (Grant No. 675451 and 823712), E-CAM (Grant No. 676531) and POP (Grant Nos. 676553 and 824080). Support from the Research Councils UK Engineering and Physical Sciences Research Council under the project `UK Consortium on Mesoscale Engineering Sciences (UKCOMES)' (Grant No. EP/R029598/1) is gratefully acknowledged. We acknowledge funding support from Research Councils UK Medical Research Council (MRC) for a Medical Bioinformatics grant (MR/L016311/1), and special funding from the UCL Provost. This work was also performed with partial support from the National Science Foundation under Grant Nos. 1562306, 1713749, 1822191, 1821431, and 1918987.\\

The authors gratefully acknowledge the Gauss Centre for Supercomputing e.V. (\url{www.gauss-centre.eu}) for funding this project by providing computing time on the GCS Supercomputer SuperMUC-NG at Leibniz Supercomputing Centre (\url{www.lrz.de}). This work used the ARCHER UK National Supercomputing Service (http://www.archer.ac.uk).



\bibliographystyle{unsrtnat_JMmod}
\bibliography{CBM2019_HemeCoupling_References}

\end{document}